# Cosmic Rays as an Interdisciplinary Earth Observation Tool: From Particle Physics and Atmospheric Processes to Geosciences and Urban Science

**Bugra Bilin[1], Nuhcan Akçit[2]**

*1. FNRS-ULB Brussels , Brussels, Belgium*

*2. Middle East Technical University Ankara, Turkey*

## ABSTRACT

The exploration of cosmic rays, which are high-energy particles originating from space and the atmosphere, has historically been associated with particle physics and astrophysics. In the last 20 years, these particles have evolved into valuable tools for observing Earth's systems. This review compiles the use of cosmic rays in three primary areas: (1) particle physics and atmospheric processes, which include cosmic-ray-induced cascades, ionization, and their impact on atmospheric chemistry and radiation; (2) geosciences, where cosmogenic radionuclides assist in the dating of geological materials and cosmic-ray neutrons are used for large-scale monitoring of soil moisture and snow water equivalents; and (3) urban science, where cosmic-ray muons are employed for non-invasive subsurface imaging and, when paired with distributed sensors, serve as the basis for smart city monitoring. This review places particular emphasis on integrating these methods with remote sensing and geographic information systems (GIS), which helps close the persistent scale gap between point measurements and satellite observations, thereby enabling three-dimensional digital representations of subsurfaces. This review concludes by discussing data standards, their integration into operational Earth observation workflows, and future research directions.

# 1.Introduction

Cosmic rays, originating from outer space, produce particle cascades that can be detected on Earth. These cascades serve as valuable tools for observing the physical phenomena they encounter along their paths, effectively acting as messengers in the universe. Since the 20th century, numerous observations of atmospheric particles have enhanced our understanding of fundamental interactions and matter. Earth observations increasingly rely on integrating various measurement types, including ground-based sensors, satellite remote sensing, airborne surveys, and numerical models. However, three fundamental challenges persist:scale mismatch between point measurements and satellite footprints [1][2], limited penetration depth of most electromagnetic remote sensing techniques, and restricted visibility of subsurface and atmospheric regions relevant to hazards and infrastructure. Cosmic rays interact with the atmosphere and the solid Earth in unique ways that complement traditional remote-sensing methods [3]. The secondary products of these interactions, such as neutrons, muons, and cosmogenic isotopes, provide valuable information on the atmospheric density and composition, near-surface water content, and subsurface density. These interactions can be quantified using ground-based detectors, cost-effective sensors, and distributed networks [4][5][6].

This review is organized into three sections. In the first section, Particle Physics and Atmospheric Issues, the interaction of cosmic rays with the atmosphere, their role in driving ionization, their influence on atmospheric chemistry and radiation, and the use of muon flux as a proxy for stratospheric temperature are explored [3][7][8]. The second section, Geosciences, discusses the application of cosmogenic nuclides and cosmic-ray neutron sensing for dating and environmental monitoring, and their connection to satellite remote sensing [4][9][10]. The third section, Urban Science, examines how muon tomography and distributed detectors facilitate infrastructure monitoring and smart city applications, and how these outputs are integrated into geographic information systems (GIS) and 3D urban models [6][11][12]. Throughout the review, emphasis is placed on how cosmic-ray-based information can be integrated with GIS and remote sensing workflows to develop coherent and multiscale representations of Earth's systems.

## 2. Particle Physics and Atmospheric Issues

### 2.1 The Basic Physics of Cosmic Rays

Cosmic rays consist primarily of bare atomic nuclei ranging from hydrogen (protons, ~90%) to iron, along with a small fraction of electrons; protons and alpha particles together account for most of the flux, while heavier nuclei up to iron contribute the remainder. These particles are accelerated in astrophysical sources such as supernova remnants and active galactic nuclei [3]. Upon entering Earth's atmosphere, these primary particles collide with nitrogen and oxygen nuclei at altitudes of approximately 15–25 km, thus initiating a cascade of secondary particles. In the context of Earth observations, three types of secondary particles are notably important: neutrons, muons, and radionuclides.

Neutrons are produced mainly through hadronic interactions and atmospheric spallation. During spallation, a high energy primary cosmic ray, or a secondary particle produced in the atmospheric cascade, collides with a nitrogen or oxygen nucleus. This interaction fragments the nucleus and releases several nucleons, predominantly neutrons, together with lighter nuclear reaction products. The flux of these particles near the Earth's surface is influenced by atmospheric pressure, humidity, and the presence of hydrogen in the ground and atmosphere [4][7]Epithermal neutrons, with kinetic energies ranging from approximately 0.5 eV to 1 MeV, are particularly relevant for cosmic-ray neutron sensors (CRNS). Hydrogen in soil water effectively moderates and absorbs these particles [4].

Muons, which are 200 times more massive than electrons and are therefore unstable, are charged particles known for their relativistic properties. They have an average lifetime of approximately 2.2 μs and travel at velocities close to the speed of light (c). These particles can penetrate hundreds of meters or even kilometers through rock, and the changes in their flux and energy during this journey provide insights into the density of the rock along their path. This ability enables the creation of density-based images [5][12][11].

Cosmogenic radionuclides are produced through two distinct pathways. Atmospheric production occurs when cosmic-ray-induced spallation reactions with nitrogen and oxygen generate nuclides such as $^{14}C$ and $^{10}Be$, which are subsequently incorporated into the carbon cycle or deposited onto the Earth's surface via precipitation and dry fallout. In-situ production, in contrast, occurs directly within minerals in rocks and sediments through spallation of target elements such as O, Si, Ca, Fe, and Cl, generating nuclides including $^{26}Al$, $^{36}Cl$, $^{21}Ne$, and, via a secondary pathway, $^{10}Be$ in quartz. Both atmospherically derived and in-situ nuclides accumulate at predictable rates, enabling their use in dating exposure and burial ages and in estimating erosion rates [13][14]. The primary cosmic-ray spectrum and flux are affected by the solar wind and geomagnetic fields, leading to variations across different latitudes, altitudes, and time periods. These variations are essential for studying the Earth's atmosphere and environment [3][15].

## 2.2 Atmospheric Cascades and Ionization

The interaction of primary cosmic rays with atmospheric nuclei results in the production of pions, kaons, and other hadrons, which subsequently decay into muons, neutrinos, and gamma rays. Neutrons can be formed in two distinct ways: through initial high-energy hadronic collisions and through subsequent spallation reactions of secondary hadrons with air nuclei within the cascade [3]. As these particles travel through the atmosphere, they electrically charge the air molecules. The greatest ionization occurs in the upper troposphere and lower stratosphere, although notable ionization is also present on Earth's surface [7]. These interactions are suspected to contribute to cloud formation through a process known as ion-induced nucleation. Ionization triggered by cosmic rays produces ion clusters that can function as condensation nuclei, possibly boosting the development of aerosol particles, and as a result, cloud droplets. This process is frequently termed the cosmic ray–cloud theory, which suggests that greater ionization during periods of reduced solar activity could lead to increased cloudiness and thus affect Earth's radiative equilibrium [16]. The CLOUD (Cosmics Leaving Outdoor Droplets) experiment at CERN investigates this hypothesis using a controlled chamber to study how galactic cosmic rays and trace vapors influence aerosol nucleation and growth under realistic atmospheric conditions [16].

Models such as Cosmic-ray Atmospheric Cascade: Application to cosmic-ray-induced ionization (CRAC:CRII v3.0) have been utilized to study space weather and atmospheric chemistry [7]. These models compute ionization levels that change with altitude, considering geomagnetic latitude, solar modulation, and atmospheric conditions. Ionization is vital for radiation exposure at high altitudes, affects the electrical properties of the atmosphere, and contributes to the creation of isotopes, such as $^{14}C$ and $^{10}Be$. These isotopes are crucial for research on ancient climates and solar activity [15][17].

In Earth observations, ionization profiles provide a physical basis for linking records of cosmogenic isotopes in ice cores and tree rings to the spatial patterns of ozone, aerosols, and radiation observed by satellites. These elements can be integrated and analyzed using GIS [17][18].

## 2.3 Atmospheric Chemistry and Ozone

Solar energetic particle (SEP) events result in the infiltration of high-energy protons and electrons into the middle and upper layers of the atmosphere. This infiltration leads to increased generation of NOx and HOx, which in turn contribute to ozone depletion, especially in polar areas [17]. Chemical climate models that incorporate cosmic-ray-induced ionization have shown that powerful SEP events, similar to the Carrington event, can lead to notable but largely temporary ozone reductions in the polar middle atmosphere [17]. In addition to natural particle-driven perturbations, emerging anthropogenic pressures on the stratospheric ozone layer are also being identified. Projections indicate that the rapid expansion of commercial spaceflight activity and the reactive species released by rocket engines could measurably delay the recovery of the ozone layer over the coming decades. This effect interacts with the polar stratospheric chemistry already sensitised by solar energetic particle events [19].

By integrating long-term cosmogenic $^{10}Be$ and $^{14}C$ records into a geographic information system (GIS) and comparing them with reanalyzed and satellite ozone data, we can reconstruct solar forcing and geomagnetic shielding over millennia. This approach offers a framework for understanding current ozone trends [15]. Recent advancements in cosmic-ray-driven electron-induced reaction (CRE) theory have further clarified the impact of ionization on halogen chemistry and ozone depletion, thereby enhancing the precision of chemistry–climate models and corroborating satellite data analysis using instruments such as the Michelson Interferometer for Passive Atmospheric Sounding (MIPAS) and Global Ozone Monitoring by Occultation of Stars (GOMOS) [20].

## 2.4 Radiation Environment for Aviation and Space flights

At standard cruising altitudes of approximately 10–12 km, cosmic rays (CRs) primarily influence the ionizing radiation environment. Effective dose rates are influenced by secondary particles, such as neutrons, protons, and muons, and these rates fluctuate based on altitude, geomagnetic latitude, and solar conditions [7]. To estimate doses along flight paths, operational models incorporate cosmic-ray fluxes, atmospheric profiles, and flight routes. Maps depicting dose rates along major air routes under varying solar and geomagnetic conditions are frequently utilized to visualize model outputs in GIS.

During ground-level enhancements (GLEs), when solar particles significantly elevate near-surface radiation, near-real-time cosmic-ray monitoring and mapping can aid in making decisions about rerouting air traffic or altering altitudes to maintain exposure within regulatory limits [6]. Additionally, smartphone-based and fixed networks, such as the Cosmic Ray Extremely Distributed Observatory (CREDO), offer another real-time data source for field radiation monitoring and may potentially detect GLEs sooner than satellite systems [6].

While the radiation environment at commercial aviation altitudes (10–12 km) is well-characterized and managed, the radiation hazard during spaceflight is orders of magnitude more severe and represents a primary limiting factor for human space exploration. In Low Earth Orbit (LEO), the International Space Station benefits from partial shielding by the Earth's geomagnetic field and residual atmosphere. However, during deep-space missions (e.g., lunar or Martian transit), astronauts are exposed to the unshielded, high-energy flux of Galactic Cosmic Rays (GCRs) and unpredictable, high-intensity Solar Energetic Particle (SEP) events [21, 22].

A critical compounding factor in spaceflight is the generation of secondary radiation. When high-energy GCRs strike the metallic hull of a spacecraft, they induce nuclear spallation, creating a cascade of secondary neutrons and light ions that can be more biologically damaging than the primary radiation itself [23]. Consequently, real-time cosmic-ray monitoring is not merely an Earth observation tool, but a vital component of space weather forecasting. Ground-based neutron and muon monitor networks provide crucial early warnings for SEP events [24], allowing mission control to instruct astronauts to seek shelter in heavily shielded storm shelters within their spacecraft. Furthermore, long-term cosmic-ray flux data are essential for accurately modeling the cumulative GCR dose, ensuring that career exposure limits for astronauts are not exceeded [25].

## 2.5 Muon Flux as an Indicator of Atmospheric Temperature

An important atmospheric application involves the use of muon flux as an indicator of stratospheric temperature. This "temperature effect" arises because mesons, such as pions and kaons, are produced at altitudes of approximately 15–20 km and can either interact with air nuclei or decay into muons and neutrinos. In a warmer and less dense stratosphere, mesons travel further before interacting, resulting in more muon decay and an increased muon flux at ground level [26][8].

To apply the temperature accurately, you must correct raw muon count rates for the full vertical atmospheric temperature profile rather than a single surface-level value. This is because different pressure layers of the stratosphere contribute unequally to meson decay probability. Special correction procedures have been created for ground-based muon hodoscopes. These procedures consider the altitude-dependent sensitivity of the hodoscopes. They use standard isobaric-level temperature data to calculate an effective temperature coefficient [27]. These correction methods are essential. They are used to isolate genuine atmospheric signals. This is from instrumental and geometric effects. This is in long-term muon monitoring records. Without such corrections, people might misattribute seasonal and short-term muon rate variations to non-atmospheric causes, undermining the reliability of muon-based temperature sounding.

Beyond the domain of temperature profiling, muon detectors function as proficient sensors, enabling the observation and monitoring of global disturbances within the Earth's magnetosphere, in addition to facilitating the prediction of space weather events. Variations in the cosmic-ray muon flux, including Forbush decreases and anisotropic flux variations occurring during geomagnetic storms, have been shown to provide critical data for understanding magnetospheric dynamics [28]. The introduction of advanced data processing techniques has recently been undertaken with a view to enhancing this capability. For instance, joint observations from the URAGAN muon hodoscope and global neutron monitor stations, processed through neural network digital analysis, have been successfully developed to predict geomagnetic storms and assess space weather hazards with high temporal resolution [29]. These developments underscore the transition of muon detectors from instruments employed in the field of pure particle physics to operational tools for space weather and geophysical monitoring.

This neural network approach processes joint observations from the URAGAN muon hodoscope and co-located neutron monitor stations, combining two independent particle detection channels into a single predictive model rather than relying on either data stream in isolation [30]. The method can capture complementary aspects of the cosmic-ray response to geomagnetic disturbances by fusing muon and neutron count-rate time series, thereby enhancing the reliability of storm onset detection in comparison with single-instrument approaches. This exemplifies a pervasive trend within the field: the integration of muon hodoscope networks with established neutron monitor infrastructure to develop sophisticated, machine-learning-driven space-weather forecasting instruments.

Both large underground and surface detectors have measured this temperature effect, which is crucial for understanding this phenomenon. The MINOS experiment determined a correlation coefficient of approximately 0.63 between the effective atmospheric temperature and seasonal variations in muon rates [8]. Similarly, the DANSS detector observed a few percent seasonal modulation of the muon flux, which was closely linked to the stratospheric temperature [31]. This makes the muon flux a valuable tool for assessing stratospheric temperature.

Networks of muon detectors can be established in the field of GIS and remote sensing using portable technologies, such as gaseous RPC detectors [32] or Micromegas [33], to derive effective temperature anomalies. A recent study [34] demonstrated that variations in the muon rate recorded at ground level directly reflect changes in atmospheric density and temperature at stratospheric altitudes. This observational approach is complemented by numerical simulation techniques, providing a quantitative framework for interpreting muon flux variations in the context of muography applications [35]. Two other recent studies [36][37] further demonstrated the possibility of cosmic-ray muons serving as natural atmospheric probes, with [36] showing that muographic imaging can monitor large-scale atmospheric structures such as tropical cyclones, and [37] introducing the muopause sounder concept, which applies the distance-of-flight muography technique to derive vertically averaged atmospheric temperature beneath the muopause . Dense urban muon networks can be utilized to study boundary layer thermodynamics and the structures of urban heat islands, complementing thermal infrared remote sensing.

In parallel with the utilisation of muography in the domains of atmospheric and magnetospheric research, the field has witnessed significant advancements in structural and geological imaging applications. In the field of cultural heritage and archaeology specifically, cosmic-ray muon radiography has been employed with considerable success to detect hidden voids and internal cavities within large artificial constructions. A seminal example is the ScanPyramids project, which utilised muon tomography to unveil a hitherto unobserved substantial void within the Great Pyramid of Khufu at Giza [38]. Similar muographic campaigns have been employed in relation to other historical and archaeological structures, including Hellenistic rock-cut chamber tombs in Naples. This demonstrates the technique's capacity for non-invasive, passive investigation of monuments where conventional drilling or geophysical survey is impractical or prohibited [39]. These heritage applications underscore the growing role of muon radiography as a diagnostic tool in conservation science and archaeological investigation.

Furthermore, muography is increasingly being explored as a tool for geological and resource prospecting. Because the technique is sensitive to integrated density along the muon trajectory, it can, in principle, delineate density contrasts associated with ore bodies, karstic cavities, and hydrocarbon-bearing formations. Recent feasibility studies have examined the use of muon radiography for imaging subsurface density anomalies relevant to mineral exploration and for constraining the geometry of geological formations in open-pit and underground mining contexts [40]. Although still in a developmental stage compared to conventional geophysical methods (gravity, seismic, electromagnetic), muographic prospecting offers the advantages of continuous passive operation, no requirement for artificial sources, and the ability to image targets at depths of several hundred metres. Integration of muographic density models with GIS-based geological maps and 3D subsurface models is expected to enhance the interpretive value of these surveys[11].

Taken together, these applications ( atmospheric temperature sounding, magnetospheric and space-weather monitoring, heritage-structure imaging, and geological prospecting) demonstrate that cosmic-ray muon detection has developed into a versatile, multi-disciplinary observational technique. Despite their differing domains, all rely on a common physical basis: the measurement of muon flux intensity or angular distribution, analysed through particle-transport physics to infer properties of the intervening medium, whether atmospheric density, magnetospheric perturbation, structural void, or subsurface density contrast. The methodological convergence of these applications is further reinforced by their increasing integration with GIS platforms, remote sensing data, and numerical modelling frameworks, which enable raw muon flux measurements to be transformed into spatially resolved, quantitatively interpretable density and structural models. This integration underscores muography's emergence as a mature, cross-disciplinary diagnostic tool applicable across atmospheric science, space physics, archaeology, and geophysical exploration.

# 3. Geosciences: Dating and Environmental Monitoring

## 3.1 Cosmogenic Radionuclides and Geological Dating

Cosmogenic nuclides generated both in the atmosphere and directly within rocks and minerals serve as essential tools for geoscientific dating [13][14]. The atmospheric production of $^{14}C$ and $^{10}Be$ underpins radiocarbon and ice-core chronologies, whereas the in-situ production of $^{26}Al$, $^{36}Cl$, $^{21}Ne$, and mineral-hosted $^{10}Be$ within rocks and sediments facilitates surface exposure and burial dating . [14].

Surface exposure dating relies on the accumulation of nuclides to determine the duration that a rock has been exposed at or near the surface and plays a crucial role in reconstructing glacier histories, changes in lake levels, and landscape evolution [14]. Burial dating, in contrast, uses the differential decay of nuclides to determine the shielding length of a material from cosmic rays. A notable example is the application of $^{26}Al$–$^{10}Be$ burial dating, which revealed that significant Australopithecus specimens in Sterkfontein Cave, South Africa, are between 3.41 and 3.7 million years old, older than previously thought [13].

Cosmogenic nuclides also enable the calculation of long-term erosion rates by examining nuclide concentrations in river sediments and bedrock, uncovering patterns of landscape erosion related to tectonics, climate, and land use [14]. Geographic information systems (GIS) and remote sensing are integral to these applications. Digital elevation models have been employed to calculate topographic shielding and altitude scaling of production rates [1][14]. Cosmogenic age maps have been combined with glacial geomorphology, lithology, satellite-derived climate, and land cover data to interpret spatial patterns [1]. Paleomagnetic excursions and changes in geomagnetic intensity observed in cosmogenic isotope anomalies can be aligned across global records using GIS-based spatial analysis [15].

### 3.2 Cosmic-Ray Neutron Sensing of Soil Moisture

Cosmic-ray neutron sensing relies on the epithermal neutron flux, which is measured just above the surface to determine soil moisture levels. The area covered by this measurement ranges from 5 to 20 ha [4][41]. Hydrogen atoms in water play a role in moderating and absorbing neutrons, implying that higher soil moisture results in a reduced neutron count [4]. Unlike conventional point-scale sensors, CRNS provides a non-invasive, area-averaged estimate of soil moisture without direct soil contact, making it particularly suitable for heterogeneous landscapes.

A stationary detector typically covers a horizontal area radius of 130–240 m and captures approximately 86% of the detected neutrons ($R_{86}$ footprint). This area expands when the soil is dry and contracts when it is wet [41], as the sensing footprint dynamically responds to changes in ambient hydrogen content (Figure 1) [42]. Vertically, the CRNS is sensitive to soil moisture within the top 15–70 cm of the soil, with a shallower effective depth under wetter conditions (Figure 2) [43]. This is significantly greater than the 2–5 cm penetration depth of most microwave remote-sensing soil moisture products [4]. Global networks such as COSMOS and COSMOS-Europe have demonstrated the scalability of CRNS deployment, showing consistent performance across diverse climate zones, land cover types, and soil textures.

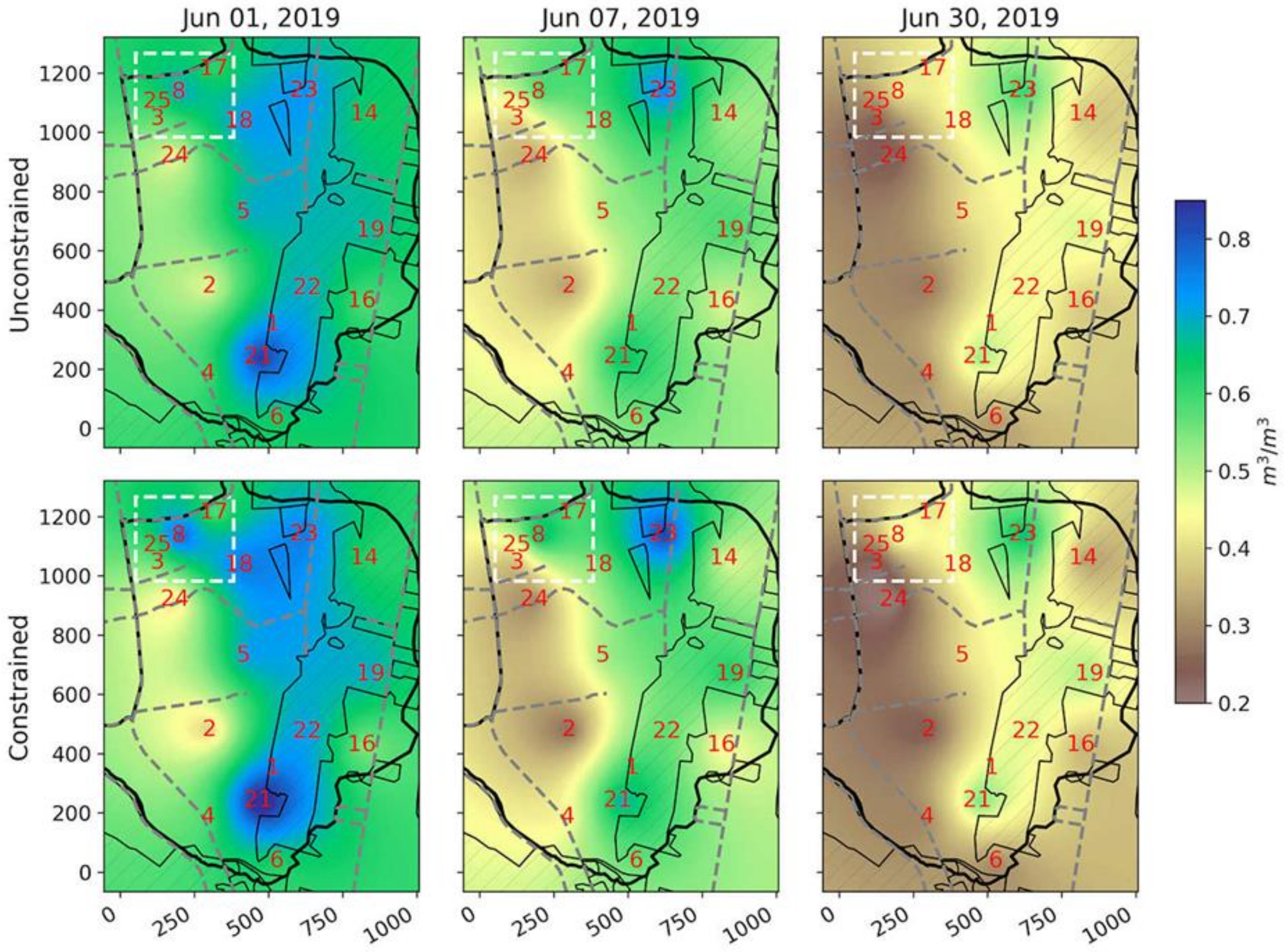

*Figure 1: Spatially interpolated soil moisture maps (m³/m³) derived from a CRNS sensor network across a study catchment on three dates in Jun. 2019, comparing unconstrained (top row) and constrained (bottom row) inversion approaches.( Heistermann et al.) [42]*

Figure 1 demonstrates why a network of CRNS instruments is valuable for hydrological Earth observation. The maps show that soil moisture is not spatially uniform within a catchment and that its spatial organization changes between wetting and drying conditions. The comparison between unconstrained and constrained inversions further shows that hydrological and landscape information can stabilize the reconstruction. Thus, CRNS networks can move beyond single-site monitoring and provide spatially explicit soil-moisture products suitable for GIS-based catchment analysis and satellite validation.

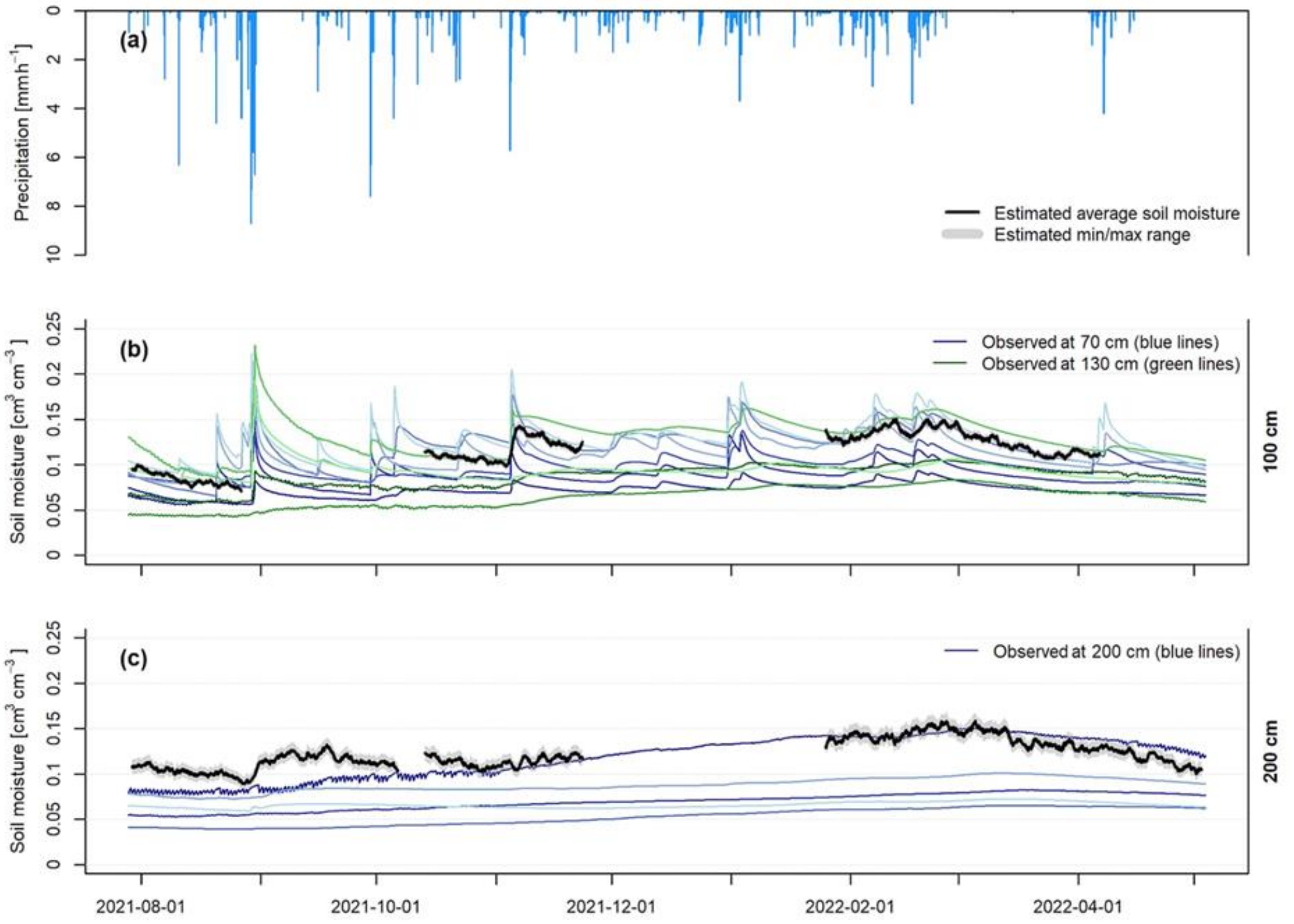

*Figure 2: Time series of CRNS-estimated soil moisture compared with in situ TDR observations across two effective sensing depth configurations from August 2021 to May 2022. (Rasche et al.) [43]*

Figure 2 emphasizes that the effective sensing depth of CRNS is not fixed. The comparison with TDR measurements at different depths shows that the CRNS signal represents a temporally varying, depth-integrated soil-water condition rather than a direct measurement at one prescribed depth. This feature is important when comparing CRNS estimates with shallow satellite retrievals and with point-based in situ probes, because each observation represents a different support volume in space and depth.

Time domain reflectometry (TDR) probes measure soil moisture within a tiny volume (~0.01 m²) immediately surrounding the probe needle, making them highly accurate but extremely localized. Satellite microwave sensors (e.g., SMAP, AMSR2, Sentinel-1) observe vast areas, from ~10 m (SAR) to ~25–50 km (passive microwave), but at the cost of spatial detail and shallow penetration depth (2–5 cm). CRNS fills the critical middle ground at ~1–20 ha, providing a spatially averaged, field-representative soil moisture estimate. This intermediate scale CRNS the ideal ground-truth tool for calibrating and validating satellite products and land surface models [41](Figure3).

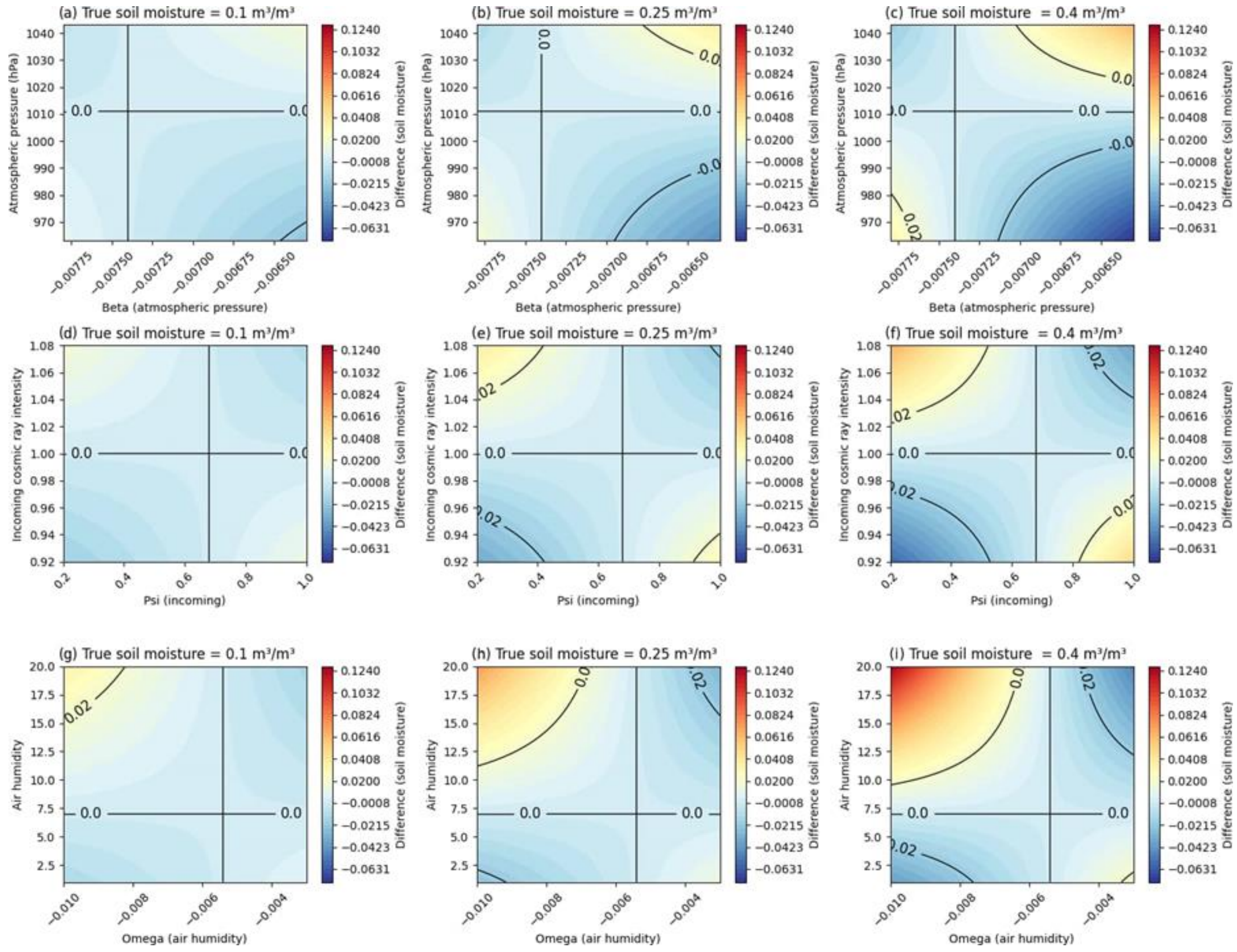

*Figure 3: Sensitivity of CRNS-derived soil moisture retrieval to key correction parameters across three true soil moisture levels (0.1, 0.25, and 0.4 m³/m³) (Baatz et al.) [41]*

Figure 3 illustrates the sensitivity of CRNS soil-moisture retrievals to correction factors related to atmospheric pressure, incoming neutron intensity, and air humidity. The magnitude and direction of the retrieval bias depend on the underlying soil-water condition, with the largest sensitivity generally occurring under wetter conditions. The figure therefore supports the need for site-specific correction, uncertainty quantification, and standardized processing before CRNS products are assimilated into hydrological models or used to validate satellite soil-moisture retrievals**.**

### 3.3 Integration with Satellite Soil Moisture Products

The application of CRNS has become standard practice for validating and enhancing soil moisture data obtained from satellite missions, such as Sentinel-1 (C-band synthetic aperture radar [SAR]) and Soil Moisture Active Passive (SMAP) (L-band radiometer). Sentinel-1 offers images with a resolution ranging from 10 to 20 m and can be revisited frequently; however, the algorithms for extracting soil moisture require training and validation data that reflect pixel-scale conditions [10]. SMAP delivers global soil moisture data, which are crucial for large-scale water monitoring [44].

Calibration using the CRNS model significantly outperformed traditional point-based calibration for Sentinel-1 soil moisture. Point-based calibration resulted in an RMSE of approximately 0.07 $m^3/m^3$ and an $R^2$ of 0.68, whereas CRNS-based calibration lowered the RMSE to approximately 0.03 $m^3/m^3$ and increased $R^2$ to 0.82, thereby reducing the error by approximately 50% [10]. This improvement is attributed to the CRNS footprint aligning more closely with the spatial averaging inherent to SAR pixels.

Similarly, researchers have employed CRNS networks to validate SMAP soil moisture products and identify systematic biases related to vegetation, surface roughness, and soil texture [44][10]. Global validation efforts have also uncovered biases in ERA5 Land soil moisture, prompting enhancements in model parameterization [10]. In GIS workflows, CRNS stations are recorded as points and linked to footprint polygons and time series; by overlaying these footprints with satellite soil moisture rasters, land cover maps, and digital elevation models (DEMs), analysts can directly compare satellite and CRNS data [10].

### 3.4 Snow Water Equivalent Monitoring

The concept of neutron moderation, which underlies CRNS soil moisture retrieval, can also be applied to estimate the snow water equivalent (SWE). Snowpacks, composed of frozen water, greatly diminish epithermal neutrons. Studies conducted in mountainous areas have demonstrated that SWE derived from CRNS can be accurate within a range of approximately ±15 mm to ±30 mm when compared with manual snow pits and high-resolution airborne lidar [9]. The CRNS is responsive to deeper snowpacks reaching depths of approximately 150 cm, exceeding the saturation range of many passive microwave satellite SWE products [9].

In environments where CRNS neutron sensitivity saturates, cosmic-ray muons offer a promising approach to SWE estimation. Muon scattering radiography (MSR) was introduced as a new technology to quantify SWE by measuring the angular deflection of cosmic-ray muons as they pass through the snowpack [45]. Continuous SWE monitoring on a Swiss alpine glacier using a muonic cosmic-ray snow gauge (μ-CRSG) was demonstrated by [46], reaching nearly 2000 mm water equivalent. Together, neutron- and muon-based cosmic-ray snow gauges provide a complementary multi-particle framework: CRNS excels for moderate seasonal snowpacks, while muon sensing extends this capability to deep glacial environments without the saturation constraints that limit passive microwave satellite retrievals.

### 3.5 GIS-Based Hydrological Applications

The incorporation of CRNS data into GIS is both a viable methodology and an increasingly operational approach in global environmental monitoring frameworks [10][1]. When CRNS-derived neutron counts are ingested into catchment-scale models via neutron forward operators, soil water storage and runoff partitioning estimates are substantially improved, with various land cover types achieving Kling–Gupta efficiencies exceeding 0.75 [47]. Mobile CRNS platforms further extend this capability by delivering spatially resolved soil moisture transects, and data fusion approaches reduce estimation uncertainty for event-scale flood and drought forecasting [48].

The COSMOS-Europe network encompasses 66 stations spread across 12 European countries, delivering harmonized, uncertainty-quantified soil moisture time series [49]. In Europe, CRNS data from ICOS and COSMOS-Europe have been added to operational flood forecasting systems, with forecasts improved by several hours in some areas [10]. At the sub-network scale, deploying a dense cluster of CRNS sensors within a single catchment reveals the spatial heterogeneity in soil moisture that wide-area networks inherently average out. A three-year observational campaign over an agricultural site in northeast Germany demonstrated that such cluster configurations substantially reduce spatial estimation uncertainty and enable high-resolution soil moisture mapping suitable for precision land-surface and agro-hydrological modelling [50] .In the US Great Plains, regular monitoring has revealed a consistent decline in soil moisture in both cultivated and native grassland areas, associated with rising temperatures, reduced groundwater extraction, and alterations in land use patterns. In Norway, evaluations of CRNS at four sites with different climate conditions have demonstrated its capability to measure rain and snowmelt in challenging high-latitude regions [51].

# 4. Urban Science: Muons, Atmosphere, GIS and Remote Sensing

## 4.1 Muon Tomography for Subsurface Imaging

Muon tomography [5][12] leverages the reduction in cosmic-ray muons to generate images of the internal density of large objects, similar to medical X-ray CT scans but using naturally occurring particles (Figure 4) [32]. As muons pass through materials such as rock, concrete, or soil, their paths are affected by the total density they encounter, thereby altering their flux and angular distributions. By measuring the muon flux from various angles and applying tomographic inversion, it is possible to reconstruct two- or three-dimensional density maps of a target.

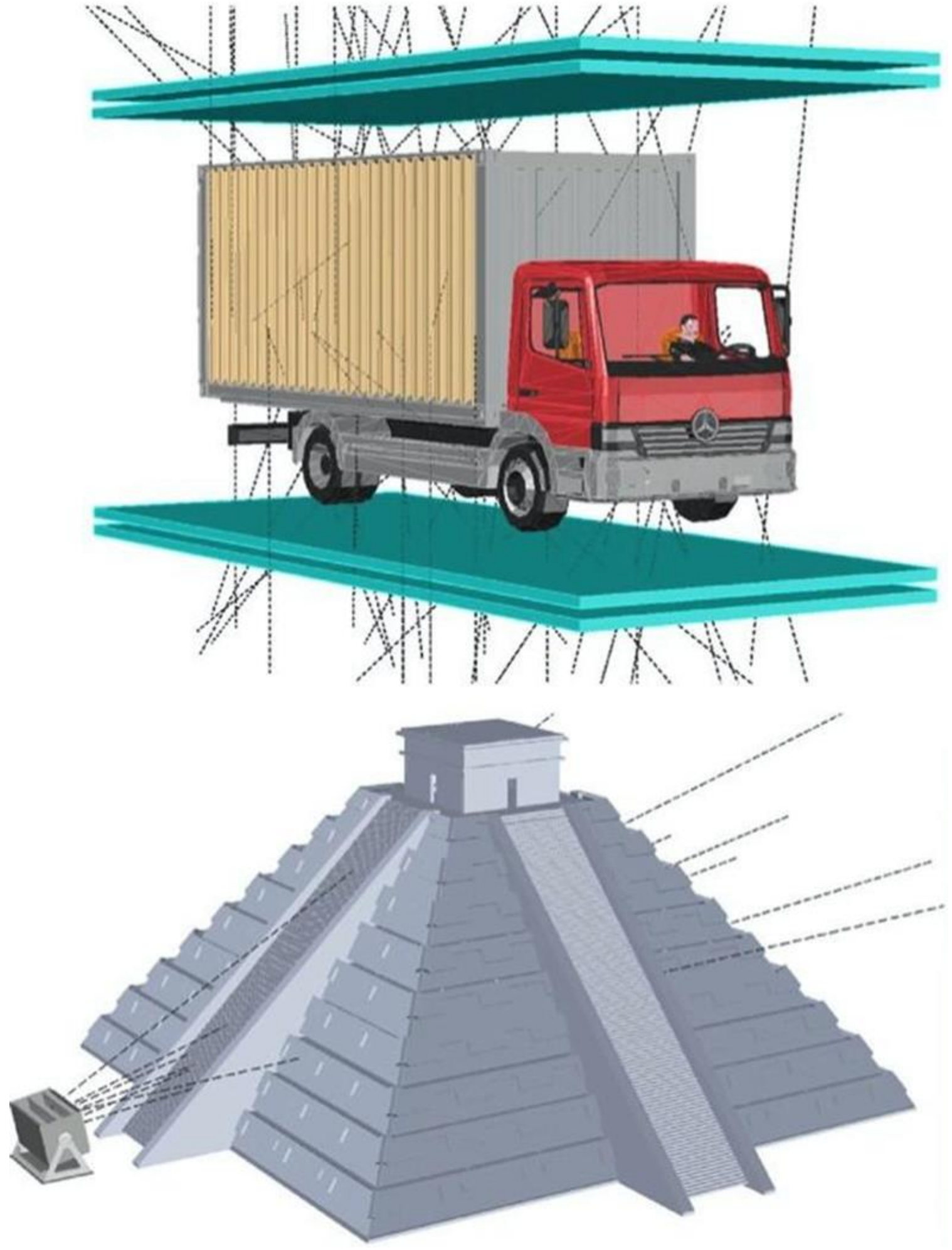

*Figure 4: Conceptual measurement geometries for cosmic-ray muon radiography. Top: transmission imaging of a cargo container, where a reduction in directional muon flux indicates higher integrated density along the muon path. Bottom: imaging of a large structure, where detectors record the angular distribution of transmitted muons to reconstruct internal density contrasts. These examples illustrate the passive, non-invasive measurement principle underlying applications to tunnels, underground infrastructure, geological voids, and heritage structures.[32]*

Figure 4 illustrates the two principal transmission-muography configurations relevant to Earth observation. In cargo or infrastructure inspection, the detector is placed beneath or behind the target and measures the directional attenuation of the incident muon flux. For large civil structures and geological targets, multiple angular views or detector positions can be combined through tomographic inversion to estimate spatial density contrasts. In both configurations, the measured quantity is not a direct image of material type but the line-of-sight integrated density, commonly expressed as density length. This distinction is important when integrating muographic results with 3D GIS, LiDAR-derived geometries, and engineering models.

In urban settings, this technology is applied to inspect underground infrastructure, such as subway systems, and evaluate the conditions of both modern and historical buildings, as well as to identify geological hazards such as karstic voids beneath cities [5][11]. A pioneering example is the work of [52], who mounted a muon telescope on a tunnel boring machine (TBM) during the excavation of a new subway line for the Grand Paris Express project, enabling real-time detection of density anomalies ahead of the advancing tunnel face. More recently, [53] performed short-duration muon flux scanning above the Xiaoying West Road subway station in Beijing, successfully resolving the overburden structure from surface-level measurements. The Shanghai Outer Ring Tunnel project [54] deployed portable dual-layer scintillator detectors inside a 3 km immersed tube tunnel beneath the Huangpu River, achieving sensitivity to sediment accumulation and real-time tidal effects. Compared with traditional geophysical methods, muon tomography offers significantly greater penetration depth (up to hundreds of meters), non-invasive operation, and continuous passive monitoring capabilities [5][11].

## 4.2 Muon Sensing for Atmospheric Temperature and GIS Integration

The measurement of muon flux can be extended beyond its use in subsurface applications, offering a distinctive integrated assessment of atmospheric structure, particularly focusing on stratospheric temperature. The positive temperature effect has been demonstrated to increase the flux of muon particles at ground or underground sites when the stratosphere experiences warming and a reduction in density [26][8][31]. Muon detectors act as ground-based temperature sounders, providing continuous data that can validate satellite microwave temperature retrievals and atmospheric reanalysis products [8][31]. Deploying a network of muon stations across a specific area allows the creation of maps showing effective stratospheric temperature anomalies for comparison with satellite and model fields within GIS software [31].

LHC experiments gather cosmic muon data during brief pauses in LHC collision data delivery, primarily to align their subdetectors (Figure 5). These data can be valuable for analyzing muon flux effects in the LHC area, matched with extreme atmospheric event information. Third, muon-based temperature monitoring techniques have proven effective in the early detection and spatial mapping of sudden stratospheric warming (SSW) events. [32] demonstrated that even a portable muon detector, deployed at the Mont Terri Underground Rock Laboratory in Switzerland, was sensitive enough to detect the abnormal atmospheric heating associated with an SSW event in January–February 2017.

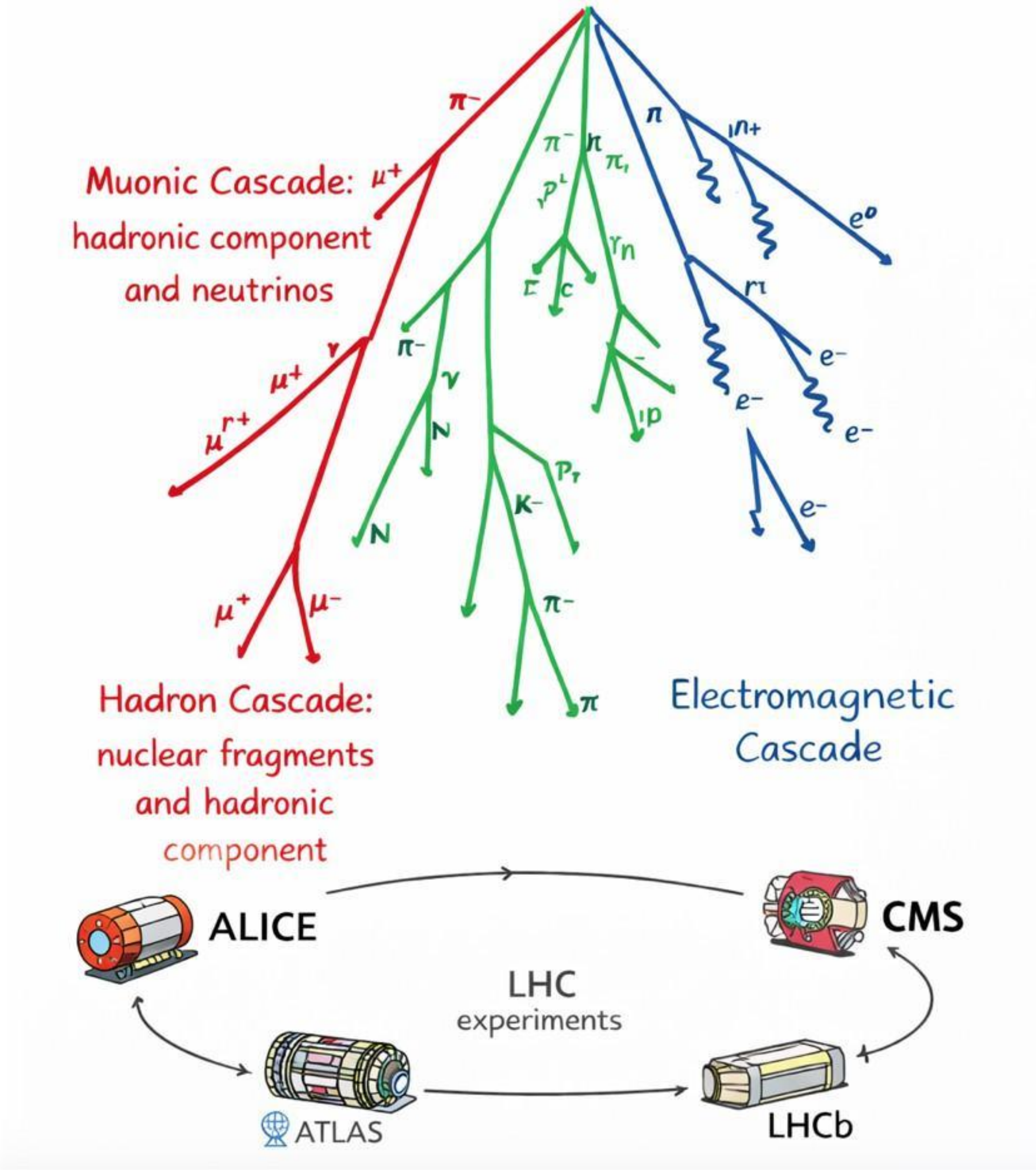

*Figure 5 : Atmospheric cosmic-ray cascade and the origin of the muon component detected at ground level by the LHC experiments. Primary cosmic rays initiate hadronic and electromagnetic cascades in the atmosphere. Charged pions and kaons may decay to muons, whose survival probability and ground-level intensity depend on atmospheric density and temperature. This dependence provides the physical basis for using muon-rate variations as an integrated probe of stratospheric thermal structure.*

Figure 5 summarizes the cascade processes that connect primary cosmic rays with the muon flux measured by ground-based and underground detectors. The atmospheric sensitivity arises mainly from the competition between meson interaction and decay. When the stratosphere is warmer and less dense, pions and kaons have a greater probability of decaying into muons before undergoing further interactions; consequently, the measured muon rate may increase. This mechanism provides the physical basis for the effective-temperature relation used in muon-based atmospheric monitoring.

Ultimately, dense urban muon detector networks have the potential to offer insights into boundary layer characteristics and urban heat island manifestations, enhancing the existing body of knowledge from thermal remote sensing and in situ meteorological observations [6].

## 4.3 Fusion with LiDAR, Photogrammetry and 3D GIS

The effectiveness of muon tomographic inversions was significantly improved by accessing high-quality three-dimensional surface geometries. Techniques such as terrestrial laser scanning (TLS) and photogrammetry using unmanned aircraft systems (UAS) can be used to generate highly detailed 3D models of structures and landscapes, allowing for the precise calculation of muon path lengths and enhancing the accuracy of density reconstruction [11].

The accurate placement of detectors, combined with ray tracing of muon paths through the surface model and iterative inversion techniques such as OSEM, aids in reconstructing voxel-based density fields [12][11]. The obtained volumetric density data can be incorporated into three-dimensional geographic information systems (3D GIS) and urban modelling platforms as voxel layers or multipatch geometries, often formatted in CityGML or similar standards [11]. This integration allows the merging of muon-derived subsurface data with surface and infrastructure datasets to support planning, maintenance, and hazard evaluation.

## 4.4 Urban Remote Sensing and Multihazard Mapping

Traditional remote sensing techniques offer a wealth of data on urban surfaces [1], such as building outlines and heights obtained from optical stereo and LiDAR, land cover and impervious surface fractions derived from multispectral imagery, land surface temperature from thermal sensors, and surface deformation from SAR interferometry. Cosmic-ray-based methods introduce additional dimensions [5][4][11][6], including subsurface density measured through muon tomography, catchment-scale soil moisture assessed by the CRNS, and atmospheric radiation and temperature patterns measured by muon networks. By integrating these multilayer datasets into GIS software, comprehensive multihazard maps can be created, offering valuable decision support for emergency management and crisis response.

For example, combining CRNS-derived soil moisture data with topography, soil type, and land use information within a GIS framework can identify areas susceptible to flooding and landslide initiation [51]. When CRNS time series indicate that root-zone moisture approaches field capacity, overlaying this information with slope gradient, drainage density, and impervious surface fraction enables the delineation of flood-prone zones with greater physical realism than static susceptibility indices alone. In parallel, a method that merges muon tomographic density maps with InSAR-derived deformation fields can aid in detecting subsidence risks and sinkhole hazards in urban settings. Integrating subsurface integrity data with transportation and utility networks is recognized as a crucial strategy for protecting critical infrastructure [11][10].

## 4.5 Distributed and Crowdsourced Cosmic-Ray Networks

A hallmark of smart city projects is the deployment of dense and affordable sensor networks. The approach to cosmic ray detection is being transformed using smartphone cameras as particle detectors, as seen in initiatives such as CREDO [6]. The CREDO project has effectively established a global, crowdsourced network for cosmic-ray detection, involving over 10,000 participants across more than 20 countries and recording over 100 million events [6]. Additionally, compact muon and neutron detectors can be mounted on structures such as buildings, streetlights, and vehicles, creating stationary and mobile urban networks [6].

A GIS is essential for the efficient management of these networks: mapping sensor locations, identifying coverage gaps, visualizing real-time anomalies in radiation and particle flux, and conducting geographic equity analyses. Current data show a notable bias towards Europe and North America, with limited observations in Africa and parts of Asia, raising concerns about the fairness of environmental monitoring [6]. During space weather events, these networks can detect GLEs in approximately 1 minute, providing early warnings to aviation and power grid operators[6].

## 5. GIS, Remote Sensing, and Data Integration

The combination of cosmic ray techniques with satellite and airborne remote sensing methods has been shown to produce more detailed depictions of urban environments [11][10]. LiDAR and photogrammetry offer precise data on surface morphology used to constrain muon path geometry. Optical and thermal images help to determine the composition and temperature of surface materials. Deformation tracking is supported by SAR, whereas CRNS gathers information on catchment-scale soil moisture and snow water equivalent (SWE). City-scale digital twins can incorporate muon-derived subsurface structures into surface building models and be further improved by adding layers for flood, heat, and subsidence risks [11][10].

Cosmic ray data appear in multiple formats, such as point time series, footprints, gridded fields, and 3D voxel models [10]. Point data are organized as point feature classes with associated attribute tables and linked time series. CRNS sensitivity regions are depicted as polygons or weighting surfaces. Gridded products are encoded as GeoTIFF or NetCDF rasters. Integrating cosmic-ray data with satellite products requires spatial overlays, temporal alignment, and statistical fusion methods (see Table 1). Multi-sensor fusion techniques, including regression and machine learning algorithms, have become essential for downscaling satellite data products [10].

Interoperability requires compliance with established standards. Open Geospatial Consortium (OGC) web services, such as sensor observation, web coverage, and web map services, facilitate the distribution of cosmic-ray data alongside other Earth observation datasets. The use of ISO 19115 metadata, CF-compliant variable names, and CityGML-based 3D representations aids in the discovery and integrated analysis [10][1].

**Table 1: Integration of Cosmic-Ray Measurements into GIS and Remote Sensing Workflows**

| Cosmic-Ray Measurement | What It Shows | GIS Integration Strategy | Combined Data Sources | Real-World Application |
|---|---|---|---|---|
| Neutrons from soil | Soil moisture content | Point with circular sensitivity zone (~200 m) | Satellite imagery, rainfall, terrain maps | Flood warnings, precision agriculture |
| Muons (through rock) | Subsurface density / voids | 3D voxel models / multipatch geometries | LiDAR laser scans, building models | Tunnel safety, structural monitoring |
| Muons (from sky) | Stratospheric temperature | Interpolated station-based maps | Weather satellites, climate models | Aviation safety, space weather monitoring |
| Crowdsourced (Phone) | Global radiation alerts | Crowd-sourced point feature classes | Flight routes, power grid data | Early warning for GLEs, solar storms |
| Simulated paths | Particle-city interaction | Vector lines through 3D city models | Urban GIS, architectural layouts | Optimization of urban detector networks |

# 6. Future Directions and Research Needs

In the field of critical research, advancing physical modelling and standardizing CRNS footprints requires a comprehensive analysis of soil mineralogy, vegetation, and topography [41]. It is crucial to expand CRNS and muon networks into areas that are currently underrepresented, particularly in the Global South, and to develop open-source software and GIS plugins for CRNS and muon tomography. A key component is shifting muon tomography from project-specific use to incorporation into the routine monitoring of critical infrastructure, along with accurately quantifying the uncertainty in data assimilation. Establishing educational and training programs to enhance capacity building is also vital [41][10][11].

Current muon measurement data from ongoing experiments can serve as a proof of concept. The phenomena affecting muon spectra can be simulated using existing tools, such as Cosmic Ray Simulations for Kascade (CORSIKA 8) [555], or specialized tools such as the Muon Simulation Code (MUSIC) and Muon Simulations Underground (MUSUN) [56]. Underground detectors can be simulated using programs such as Geometry and Tracking (GEANT4) [57] and DELPHES [58]. These simulation data, when combined with other direct probes, can provide a precise model of the observability of such phenomena (see Table 2). Recent pocket-sized detector technologies, such as a portable Arduino-based muon detector [59] and the CosmicWatch v3X [60], have extended cosmic-ray measurements to educational and citizen-science settings.

**Table 2: Key Simulation Tools for Cosmic-Ray Earth Observation**

| Tool | Core Purpose | Primary Application Example |
|---|---|---|
| CORSIKA 8 | Atmospheric cascade modeling | Generating ionization profiles and spectra |
| GEANT4 | Detector response simulation | Simulating muon flux in complex urban canyons |
| MUSIC / MUSUN | Muon transport in rock/subsurface | Forward modelling for muon tomography |
| DELPHES | Fast detector simulation | Rapid parametric simulation of detector response for muon flux measurements in large-scale experiments |

The European strategic research investment plan for cosmic ray Earth observation (2025–2035) envisions a coordinated investment in international standards, open-source tools, network expansion, and operational integration. An estimated budget of approximately €345 million has been proposed for standardization, network deployment, capacity building, integration of forecasting systems, GEOSS catalogue integration, and advanced detector research and development [10][1].

## Conclusion

Initially focused on particle physics and astrophysics, cosmic ray research has expanded into a multidisciplinary field with important applications in Earth observation, impacting atmospheric studies, geosciences, and urban analysis [3][4][5][11]. Cosmic rays contribute to ionization and affect atmospheric chemistry and radiation. Muon flux serves as a valuable ground-based indicator of stratospheric temperature, enhancing data from satellite sounders and re-analyses [7][15][8][22][20].

In geosciences, cosmogenic nuclides are used as chronometers to explore landscape changes and human history [13][14]. Moreover, CRNS is crucial for assessing soil moisture and snow water equivalent at the ideal scale for validating satellite data and calibrating hydrological models [4][9][10][44]. In urban science, muon tomography has been applied to detect subsurface structures with sub-meter precision [5][12][11][6], and distributed detector networks have been employed for radiation and hazard monitoring.

Geographic information systems and remote sensing methods offer a framework that enables the integration of various data streams, supporting footprint mapping, satellite and in situ data fusion, detailed 2D and 3D environmental modelling, and comprehensive multihazard analyses [1][10][11]. By strategically investing in standards, networks, and capacity development, cosmic-ray Earth observations can become a key component of global monitoring systems, significantly improving the global ability to manage future water resources, infrastructure, and climate-related risks.

## Acknowledgements

The authors would like to express their sincere gratitude to Dr. Andrea Giammanco for his thorough proof reading and for providing detailed comments that greatly helped improve this review article.